\documentstyle[aps,prl,multicol]{revtex}
\begin{document}
\title{\bf Persistence in a Stationary Time-series}

\author{Satya N. Majumdar$^{(1),(2)}$ and Deepak Dhar$^{(1)}$}

\address{(1) Tata Institute of Fundamental Research, Homi Bhabha Road,
Mumbai-400005, India. \\
(2) Laboratoire de Physique Quantique (UMR C5626 du CNRS),
Universit\'e Paul Sabatier, 31062 Toulouse Cedex, France. \\}

\date{\today}

\maketitle

\begin{abstract} 
We study the persistence in a class of continuous stochastic processes 
that are stationary only under integer shifts of time. We show
that under certain conditions, the persistence of such a continuous process
reduces to the persistence of a corresponding discrete sequence obtained
from the measurement of the process only at integer times.
We then construct a specific sequence for which the persistence can be
computed even though the sequence is non-Markovian.
We show that this may be considered as a limiting
case of persistence in the diffusion process on a hierarchical lattice.

\noindent

\end{abstract}

\begin{multicols}{2}

\section{Introduction}

In recent years, there has been a lot of interest in the study of
persistence of fluctuations in different physical systems
\cite{Review,Rev2}. Persistence $P(t)$ is simply the probability
that the deviation of the value of a
fluctuating field from its mean value does not change sign up to time $t$.
Persistence has been studied in many
nonequilibrium systems\cite{Review} and also in diverse fields ranging
from ecology\cite{Ecology} to seismology\cite{Seismology}. 
Theoretical studies include various models of
phase ordering kinetics\cite{Po}, diffusion equation\cite{Diffusion,D2},
reaction diffusion systems in pure\cite{Rd} as well as
disordered environments\cite{Drd}, fluctuating interfaces\cite{Interfaces},
and various theoretical models\cite{Sp}. Persistence or first
passage properties find simple applications in various
chemical\cite{Shapir},
biological\cite{Knots} and granular systems\cite{Gran}.
In laboratory experiments, persistence has been measured in various experimental systems
including breath figures\cite{Marcos}, liquid crystals\cite{Yurke}, soap
bubbles\cite{Tam} and laser-polarized Xe gas using NMR
techniques\cite{Wong}.

In many of the nonequilibrium systems discussed above, the underlying
stochastic process $\psi(t)$ is nonstationary. For example, the two time
correlation function $C(t_1,t_2)=\langle \psi(t_1)\psi(t_2)\rangle$ for
the diffusion equation depends on the ratio of the two times $t_1$ and
$t_2$, and not on their difference\cite{Diffusion}. Persistence in such
systems typically decays as a
power law $P(t)\sim t^{-\theta}$ at late times $t$. The exponent $\theta$,
called the persistence exponent, is believed to be a new exponent and is
apparently unrelated to the usual dynamical exponents that characterize
the decay of $n$-point correlation functions with finite $n$.  
Persistence has also been studied for stationary
processes\cite{BL,Slepian} such as a  stationary Gaussian process
characterized by its two time correlation  $C(t_1,t_2)$ which is invariant
under {\it
arbitrary} time translation, i.e., $C(t_1+t_0,t_2+t_0 )=C(t_1,t_2)$ for
all $t_0$. In the stationary case, persistence between times $t_1$ and
$t_2$ typically decays exponentially, $P(t_1,t_2)\sim
\exp[-\theta_s|t_2-t_1|]$ for large time difference\cite{Slepian}. For
some processes such as the diffusion equation the nonstationary problem
can be mapped onto a corresponding stationary one\cite{Diffusion} and the
exponent $\theta$ of the nonstationary process becomes identical to the
inverse decay rate $\theta_s$ of the corresponding stationary
process\cite{Review}. Despite many theoretical studies of either $\theta$
or $\theta_s$, exact results are known only in relatively few
cases\cite{Exact}. The basic difficulty in computing either of them can be
traced back to the fact that the underlying stochastic processes in both
cases are usually non-Markovian\cite{Review}.

In this paper we study the persistence in stochastic processes that are
stationary under translations in time only by an {\it integer multiple of
a basic period} (without loss of generality, this period may be chosen to
be $1$). Throughout this paper we will refer to such processes as SIS
(stationary under integer shifts). For example, a Gaussian stochastic
process will have the SIS property if its two-time 
correlation function $C(t_1,t_2)$ satisfies
$C(t_1+n,t_2+n)=C(t_1,t_2)$
for all integer $n$. Such processes appear in many physical
situations.
For example, in weather records, there is an underlying non-random
periodic forcing ( the motion of earth round the sun), which makes the
stochastic process not truly stationary
in time.  In nonlinear systems, even if one can filter out the periodic
component, the properties of the filtered signal (say variance) would
still be expected to show a periodic variation with time. It seems
worthwhile to study in more detail persistence in such SIS processes.

When one wants to study the persistence of SIS processes, the following
question arises naturally: Is the probability $P(t)$ that the process
remains positive over the interval $[0,t]$ is same as the probability
$P_n$ that the process is positive only at all the $n$ intermediate
integer times between $0$ and $t$?  In other words, is the persistence of
a `continuous' SIS process is the same as the persistence of the
corresponding `discrete' sequence obtained by measuring the process only
at integer times?

The question regarding the difference between `continuous-time' and
`discrete-time' persistence was first raised in Ref. \cite{MBE} for
strictly stationary Gaussian processes, motivated from the observation
that in experiment\cite{Wong} as well as numerical simulation\cite{NL} of
persistence in diffusion equation, the measurements (whether the process
is positive) are done only at discrete points (separated by a fixed time
window of size $\Delta$) even though the actual process is continuous. In
Ref. \cite{MBE} it was shown that for general stationary Gaussian
processes, the continuous persistence decays as 
$P(t)\sim \exp[-\theta_c t]$ for large $t$, where as the 
corresponding
`discrete-time' persistence (obtained from measuring the data only at the
intermediate time points separated by a fixed $\Delta$) decays as $P_n
\sim \exp[-\theta_d n]$ where $T=n\Delta$.  In general, one would expect
that the exponent $\theta_c$ is strictly greater than $\theta_d$ for such
a process since the process can change sign between two successive integer
points.  The exponent $\theta_d$ was computed analytically in Ref.
\cite{MBE} for a Gaussian stationary Markov process and was shown to
depend continuously on the window size $\Delta$.

In this paper we study the `continuous-time' versus `discrete-time' persistence for
SIS processes. We restrict ourselves to the study of discrete-time persistence
only when the measurement points in time are integers. This is a natural
choice since the process is stationary only under integer shifts in time.
We show that unlike strictly stationary processes, in this case, the two
exponents $\theta_c$ and $\theta_d$ can be equal under certain conditions.
This discrete-time persistence $P_n$ of a sequence is also rather interesting
from a purely mathematical point of view, especially when the underlying
process $\psi(t)$ is Gaussian. In that case, calculating $P_n$ becomes the
problem of calculating the probability that a set of $n$ Gaussian random
variables with a specified joint probability distribution are all
positive. This `one-sided barrier' problem has remained popular in the
applied mathematics literature for many decades\cite{BL,Slepian,Gupta}.  
But the number of cases where this probability can be explicitly
calculated for large $n$ remains rather small\cite{Gupta}.

In this paper, we study a specially simple case of a continuous-time
stochastic process $\psi(t)$ which is obtained as a local smearing of a
sequence of independent identically distributed random variables via a 
smearing function $f(t)$. This
process becomes, by construction, a stochastic process whose probability
distribution is invariant under time translations by integers, i.e. a SIS
process. We construct examples where $\theta_d < \theta_c$, and also
construct a family of smearing functions $f(t)$ for which $\theta_d =
\theta_c$. We provide a physical example namely the diffusion equation on
a hierarchical lattice where the diffusion field is a Gaussian stochastic
process with the SIS property and we compute the corresponding smearing
function exactly. We then determine exactly the exponent $\theta_d$ for a
specific case when the correlations in the discrete sequence are nonzero
only for consecutive values. We find that in this case, the exponent
$\theta_d$ depends continuously on the value of the correlation.

The paper is organized as follows. In section II, we give some examples of
continuous SIS processes where $\theta_c $ is strictly greater than
$\theta_d$, and also construct a class of processes for which
$\theta_c=\theta_d$. In section III, we provide a physical example namely
diffusion equation on a hierarchical lattice where the diffusion process
shows log-periodic oscillations. After rescaling, and a change of
variables from time $t$ to $log(t)$, we get a stochastic process that has
the SIS property. In section-III, we introduce a special sequence for
which the persistence exponent $\theta_d$ can be computed exactly.  
Section IV contains a summary of our results.

\section{Continuous-time versus Discrete-time Persistence}

In this section we discuss the conditions under which the continuous-time
persistence of a SIS process is the same as the discrete-time persistence of a
corresponding sequence obtained by measuring the process only at the
integer points. As noted earlier, in general, we expect that $\theta_c >
\theta_d$. Consider a
stochastic process $\psi(t)$, which is known to be positive at the integer
points $t=1,2, \ldots N$. Now consider the conditional distribution of
$\psi(t_1)$ at some non-integer point $t_1$ lying within the interval
$[1,N]$. This conditional distribution is a gaussian whose width is
independent of the values of $\psi(t)$ at the known integer points. If
this variance is finite, even in the limit when $N$ is large, clearly,
there will be finite probability of $\psi(t_1)$ becoming negative. Thus
one would expect to get $\theta_c = \theta_d$ only if the conditional
variance of $\psi(t_1)$ tends to zero for large $N$. In such a process,
one should be able to determine $\psi(t)$ for all real $t$, if one knows
its value at all integer points.

This suggests the following construction of $\psi(t)$ : we consider a
sequence of independent random variables
$\{\phi(n)\}$ having zero mean, where $n \in
\{-\infty,+\infty\}$, and define a stochastic process $\psi(t)$
by the convolution
\begin{equation}
\label{defpsi}
\psi(t) = \sum_{n=-\infty}^{+\infty} f(t-n) \phi(n).
\end{equation}
Knowing $\psi(t)$ at all integer points, one can expect to determine 
uniquely the constants $\{\phi(n)\}$ by solving coupled linear equations,
which then determines $\psi(t)$ for all real values of $t$.

The behavior of this process depends only on the smearing function $f(t)$.
In the following, we shall assume that $f(t)$ has some good   
properties, i.e. is a non-negative unimodal function of $t$, which
decreases sufficiently fast for large $|t|$.  By a shift of the origin of
time $t$, and rescaling $\psi(t)$, we can assume that the maximum of
$f(t)$ occurs at $t=0$ and $f(0)=1$.

What is the class of functions $f(t)$ such that  $\theta_c$ equals
$\theta_d$? This class is not easy to characterize directly. A simple
example illustrates this point clearly. Consider the simple case of
triangular function
\begin{eqnarray}
f(t) & = & 1 -|t|/a  , \text{~~for}~~ |t| < a
\\
     & = & 0, \text{otherwise.}
\end{eqnarray}

In this case, $\psi(t)$ is a piece-wise
linear function of $t$. If $a < 1/2$,  we have intervals in which
$\psi(t)$ is identically zero.
If, however, we define persistence probability as the probability that the
function does not {\it change sign} upto time $t$, it is clear that for
all $a < 1$, we have $\theta_c =  \theta_d = log 2$.

We now show that $\theta_c \neq \theta_d$ if $a > 1$. For this purpose, it
is sufficient to show that there are sequences $\{\phi_n\}$ such that the
corresponding $\psi$-process is positive at all integer points,  but takes
negative value for non-integer $t$.  As such events would occur with non-zero
probability along the sequence, $\theta_c > \theta_d$. 

Let $n$ be the integer just below $a$. We consider a periodic sequence
of $\phi_n$ with 
\begin{eqnarray}
\phi_i &=& 1, \mbox{ if $i=0$ mod$(2 n +1)$}; \nonumber \\
       &=& -c, \mbox{ if $i= n$ or $n+1$ mod $( 2 n +1)$};\nonumber
\\
       &=& 0 \mbox{ otherwise.}
\end{eqnarray}
Then it is easy to see that if we choose $c$ such that 
$( a - n) > ( 2 a - 1) c > ( a -n - 1/2)$, then $\psi(t=i)$ is
positive for all integers $i$, but $\psi(t=n+1/2)$ is negative. Clearly,
these signs are not changed if all $\phi$'s deviate from these values by
sufficiently small amount. Then such sequences ( of finite length) will
occur with non-zero frequency, and hence for any $a > 1$, $\theta_c$ is
strictly greater than $\theta_d$.

However, there are functions $f(t)$ for which   $\theta_c = \theta_d$.  
The simplest example of this class is  $f(t) =
\exp( -|t|/a)$. In this case, it is easy to see from Eq. (\ref{defpsi}) 
that $\psi(t)$ at any 
non-integer point $t$ can be expressed as a positive linear combination of
its value at the two nearest integer points, so that for all $t = n +
\delta t$ with 
$0 \leq \delta t \leq 1$ we have
\begin{eqnarray}
\label{sinh}
\psi(n + \delta t) &=&[ \sinh \frac{\delta t }{a} \psi(n+1)\\
&+&\sinh\frac{(1-\delta t)}{a}
\psi(n)]/\left[\sinh(1/a)\right]
\end{eqnarray}
Thus, if $\psi(n)$ and
$\psi(n+1)$ are positive, Eq. (\ref{sinh}) implies that $\psi(t)$ is positive for all
$n \leq t \leq n+1$. Hence one gets, $\theta_c = \theta_d$.

The example above can be generalized. For example, one can introduce a two parameter family of 
functions, $f(t)= \exp(
\kappa_1 t)$ for $t<0$, and $f(t) = \exp(-\kappa_2 t)$ for $t > 0$ with
$\kappa_1>0$ and $\kappa_2>0$ and are not equal in general. In
fact, one can even introduce two arbitrary periodic functions $g_1(t)$ and
$g_2(t)$ ( with period $1$), and take
\begin{eqnarray}
f(t) & = & \exp(  \kappa_1 t - g_1(t)) , \text{~~for} ~~t < 0,
\\
     & = & \exp( - \kappa_2 t - g_2(t)) , \text{~~for} ~~t > 0
\end{eqnarray}
without destroying the equality of $\theta_c$ and $\theta_d$. One only has
to impose some conditions on $g_1(t)$ and $g_2(t)$ to ensure that $f(t)$ is
unimodal. Effectively, we can take any unimodal function $f(t)$ defined
in the interval $-1 \leq t \leq 1$, and extend it to all real $t$ using
the conditions $f(t-1) = e^{ - \kappa_1} f(t)$, for $t <0$, and 
$f(t+1) = e^{ -\kappa_2} f(t)$ for $t > 0$, to get a function $f(t)$ for
which $\theta_d$ and $\theta_c$ are equal. 
 
\section{Persistence in Diffusion Equation on a Hierarchical Lattice}

A simple example of a physics problem where functions of the type
given by Eq.(\ref{defpsi}) show up is the  persistence of a
diffusion field on a hierarchical lattice.
The lattice may be thought of as a line having $N=2^n$ sites
labelled by an n-bit binary integer $ i, 0 \leq i \leq N-1 $\cite{dyson,collet}.
We define
the ultrametric distance between two sites $i$ and $j$ as $d$, if the
binary integers denoting $i$ and $j$ differ at the $n-d+1$ bit counting
from the left. Thus we have $d=1$  between sites $2$ and $3$, but $d=4$
between sites $7$ and $8$. ,
At each site $i$, we have a real variable $\psi(i)$. At time $t=0$,
the fields at different sites are assumed to be independent identically
distributed random variables (say gaussians of mean zero, and variance 1).
The fields ${\psi(i)}$ are assumed to evolve in time by the deterministic
equation
\begin{equation} 
\frac{d}{dt} \psi(i) = \sum_{j=0}^{N-1} K_{i,j} [ \psi(j) - \psi(i)]
\end{equation}

Here the spring constants $K_{i,j}$ are assumed to be functions of the   
distance $d_{i,j}$ between the two points. In the following, we shall
assume that $K_{i,j} = a^{ -  d_{i,j}}$, where $a$ is a constant $ > 1$.

The integration of the equations of evolution is made particularly simple
by the
hierarchical nature of the spring couplings.  It is easily verified that 
we have $2^{n-r}$ independent
eigenmodes of relaxation rate $( a -1)^{-1} a^{-r +1}$ ($r = 1,2,\ldots
N-1$)
satisfying
\begin{equation}
\frac{d}{dt} S_j^{(r)} = (a -1)^{-1} a^{-r+1} S_j^{(r)}
\end{equation}
where
\begin{equation} 
S_j^{(r)} = \sum_{k=0}^{2^{r-1}-1} \phi(j 2^r -2^{r-1}-k) - \phi(j 2^r -
k)
\end{equation}
where $j = 1$ to $ 2^{n-r}$.

Expanding any particular $\psi(i)$, say for $i=1$, in terms of these
eigenvectors, and we get,
\begin{equation}
\psi_1 (t)  = \sum_{r=1}^{n-1} 2^{-r/2} \exp\left[ - (a -1)^{-1} a^{-r+1} t\right]
\phi(r),
\end{equation}
where $\phi(r)$'s are i.i.d. Gaussian variables of zero mean and unit
variance that characterize the initial condition.
This formula for the hierarchical model may be compared with the
corresponding formula one writes in the Euclidean space in $d-$dimensions
\begin{equation}
\psi(\vec r=0) = \int_0^{\infty}  dk \exp( - k^2 t) \eta(k)
\end{equation}
where $\eta(k)$ are  white-noise process with variance
\begin{equation}
< \eta(k) \eta(k')> = \delta_{k,k'} k^{d-1}
\end{equation}
We eliminate the time variable $t$ in terms a logarithmic time variable 
$\tau$ using the identification $ a^{\tau} = t$, and change $\psi(t)$
by a change of scale, 
$\psi(\tau) = [a^{\tau/2}]  \psi(t = ( a-1)  a^{\tau})$.
Then we have
\begin{equation} 
\psi(\tau)= \sum_{r=0}^{\infty} \xi(r)\exp \left[- a^{\tau-r}\right]  a^{
(\tau-r)/2}.
\end{equation}
For large $\tau$, the summation over $r$ can be extended from $-\infty$ to
$+\infty$, and the process $\psi(\tau)$ then becomes a Gaussian process
with the SIS property, i.e., is stationary only under integer shifts in time
and is obtained by local smearing of the discrete white noises $\phi(r)$'s,
\begin{equation}
\psi(\tau) = \sum_{r=-\infty}^{+\infty} f(\tau - r) \phi(r),
\label{Eq:psi}
\end{equation}
where the convolution function $f(r)$ clearly goes to zero when $r$ tends
to $\pm \infty$.
Thus, the problem of calculating the persistence exponent reduces that of
calculating the exponents $\theta_c$ and $\theta_d$ for a process defined
by given convolution
function $f(t) = \exp(-a^t) a^{t/2}$. The origin of the SIS property here
comes from the discrete scale invariance of the original model, which
gives rise to log-periodic oscillations in the relaxation processes
\cite{sornette}.

We have not been able to determine whether this function $f(t)$, the
exponents $\theta_c$ and $\theta_d$ coincide, or are different. However,
in a simple Monte Carlo realization of a sequence of $10^5$
Gaussian variables $\{\phi_i\}$'s of zero mean and unit variance,
we did not find any
instance where the function $\psi(t)$ changed sign twice between two
consecutive integers. This indicates that these exponents, if not equal,
are likely to be quite close to each other.

\section{Exact Results for a Special Case}

For a smearing function $f(t)$ for which $\theta_c=\theta_d$, the
computation of the persistence exponent simplifies considerably, and
reduces to its determination for a discrete sequence
rather than a continuous process. But even then, the exponent $\theta_d$
is quite nontrivial for an arbitrary smearing function $f(t)$.
For calculating $\theta_d$, only the values of $f(t)$ at
integer points are relevant. In the following, we shall 
consider in detail the
calculation of $\theta_d$ when only $f(0)$ and $f(-1)$ are non-zero. 
This can be thought of a crude approximation to the smearing function 
$f(t) = \exp(-a^t) a^{t/2}$, as in the diffusion equation on a hierarchical lattice,
which decreases superexponentially 
for $t>0$ and only exponentially for $t<0$ for $a>1$. We will show below 
that the exact computation of $\theta_d$ is nontrivial even for this
toy smearing function since the resulting sequence is non-Markovian.  

In this special
case, Eq. (\ref{Eq:psi}) becomes
\begin{equation}
\psi_i = \phi_i + \epsilon\, \phi_{i+1},\,\,\,\,\, i=$1$,$2$,$\ldots$,$n$
\label{psi1}
\end{equation}
where we shall assume that $\{\phi_i\}$ are independent identically distributed random
variables, not necessarily Gaussian, each drawn from the same distribution
$\rho(\phi)$. Here $\epsilon$ is
a mixing parameter. For convenience, we relabel the 
$\phi$'s without any loss of generality
to consider the following sequence,
\begin{equation}
\psi_i = \phi_i + \epsilon\, \phi_{i-1},\,\,\,\,\, i=$1$,$2$,$\ldots$,$n$.
\label{psi}
\end{equation} 
For 
simplicity, we will assume that $\rho(\phi)$ is
symmetric about the origin. The mean value of $\phi$ is then zero. We now
ask: what is the
probability $P_n(\epsilon)$ that 
$\psi_1, \psi_2, \ldots \psi_n$  are all  positive for a given $\epsilon$?

We note that the variables $\psi_i$'s are now correlated. The two point
correlation function, $C_{i,j}=\langle \psi_i \psi_j\rangle$ can be easily computed
from Eq. (\ref{psi}),
\begin{equation}
C_{i,j}= {\sigma}^2\left[ (1+\epsilon^2)\delta_{i,j}+\epsilon
(\delta_{i-1,j}+\delta_{i,j-1})\right],
\label{corr}
\end{equation}
where $\delta_{i,j}$ is the Kronecker delta function and
$\sigma^2=\int_{-\infty}^{\infty}\phi^2\rho(\phi)d\phi$.
Thus the parameter $\epsilon$ serves as a measure of the correlation
and it is this correlation that makes
the calculation of $P_n(\epsilon)$ nontrivial for nonzero $\epsilon$.

The sequence $\{\psi_n\}$ defined by Eq. (\ref{psi}) is non-Markovian
in the sense that if only $\{\psi_n\}$ are observed, and not the
$\phi_n$'s,  $\psi_n$ depends
not just on the previous member of the sequence $\psi_{n-1}$, but rather
on the whole history of the sequence. For example, from Eq. (\ref{psi}) 
one can express $\psi_n$ as,
\begin{equation}
\psi_n=\sum_{k=0}^{n-1}(-1)^{k-1}{\epsilon}^{k}\psi_{n-k} +\phi_n-{\epsilon}^n\phi_0
\end{equation}
which demonstrates explicitly the history dependence of the sequence. For non-Markovian 
sequences, it is generally hard to compute the persistence exponent.
Fortunately progress can be made for this special sequence even though it is non-Markovian.

In order to calculate $P_n(\epsilon)$, it is first useful to define
the following probabilities,
\begin{eqnarray}
Q_1(x)&=&\int_x^{\infty} d\phi_0 \rho(\phi_0), \nonumber \\
Q_n(x)&=&\int_x^{\infty}d\phi_0 \rho(\phi_0)\int_{-\epsilon \phi_0}^{\infty}d\phi_1 \rho(\phi_1)
\int_{-\epsilon \phi_1}^{\infty}d\phi_2 \rho(\phi_2)\ldots \nonumber \\
&&\ldots \int_{-\epsilon \phi_{n-2}}^{\infty}d\phi_{n-1} \rho(\phi_{n-1}),\,\,\,\,\,\, n\geq 2.
\label{qnx}
\end{eqnarray}
Using the definitions in Eq. (\ref{psi}), it is then easy to see that the persistence
$P_n(\epsilon)=Q_{n+1}(-\infty)$. This is due to the fact that for all the $\psi_i$'s
in Eq. (\ref{psi}) to be positive, while $\phi_0$ is
free to take any value, $\phi_1$ must be bigger than $-\epsilon \phi_0$,
$\phi_2$ must be bigger than $-\epsilon \phi_1$ and so on.
Differentiating Eq. (\ref{qnx}) with respect to $x$, we get the recursion relation
\begin{equation}
{ {d Q_n(x)}\over {dx}}= -\rho(x) Q_{n-1}(-\epsilon x), \,\,\,\, n\geq 1,
\label{recur1}
\end{equation}
with $Q_0(x)=1$ and the boundary condition, $Q_n(\infty)=0$ for all $n\geq 1$.
Let us define the generating function
\begin{equation}
F(x,z)=\sum_{n=1}^{\infty}Q_n(x)z^n.
\label{gen}
\end{equation}
>From Eq. (\ref{recur1}), it follows that $F(x,z)$ satisfies a first
order {\it non-local} differential equation,
\begin{equation}
{ {\partial F(x,z)}\over {\partial x}}=-\rho(x)z\left[1+F(-\epsilon x, z)\right],
\label{diff}
\end{equation}
with the boundary condition, $F(\infty,z)=0$ for any $z$. Once we know the function
$F(x,z)$, $P_n(\epsilon)$ can be obtained by evaluating the Cauchy integral,
\begin{equation}
P_n(\epsilon)=Q_{n+1}(-\infty)={1\over {2\pi i}}\int_{C_0}{ {F(-\infty,z)}\over {z^{n+2}}}dz,
\label{cauchy}
\end{equation}
over a contour $C_0$ encircling the origin in the complex $z$ plane.

Before proceeding to solve Eq. (\ref{diff}), we make the simple observation that,
\begin{equation}
P_n(\epsilon)=P_n\left({ {1}\over {\epsilon}}\right),
\label{symm}
\end{equation}
true for any $\epsilon$.  To see this, we first rescale the $\psi_i$ variables,
${\psi_i}'= \psi_i/{\epsilon}$. Clearly the persistence of ${\psi_i}'$'s is the
same as that of the $\psi_i$'s. Dividing Eq. (\ref{psi}) by $\epsilon$, we see that
in order for the ${\psi_i}'$'s to be positive, we need to satisfy the
conditions: $\phi_0>-\phi_1/\epsilon$, $\phi_1>-\phi_2/\epsilon$, $\ldots$,
$\phi_{n-1}>-{\phi_n}/\epsilon$ where $\phi_n$ can be arbitrary. Eq. (\ref{symm}) then
follows once we relabel
$\phi_i \to \phi_{n-i}$ for all $0\leq i\leq n$. Thus it is sufficient to compute
$P_n(\epsilon)$ for $\epsilon$ only in the range, $-1\leq \epsilon \leq 1$. Once we know
this,
$P_n(\epsilon)$ for $|\epsilon|>1$ can be obtained from Eq. (\ref{symm}).

Let us summarize our main results. We show that for $-1< \epsilon \leq 1$,
$P_n(\epsilon)\sim
\exp (-\theta (\epsilon) n)$ for large $n$, where $\theta(\epsilon)$
depends continuously on $\epsilon$ and also depends on the distribution
$\rho(\phi)$. In contrast, at $\epsilon=1$, the exponent
$\theta(1)=\log ({{\pi}\over {2}})$ is  independent
of the distribution $\rho(\phi)$. The
exponent $\theta(\epsilon)$ diverges as $\epsilon\to -1$, indicating
a faster than exponential decay of $P_n$ for large $n$.
We show that $P_n(-1)=1/(n+1)!$ exactly for all $n\geq 1$,
again independent of the distribution $\rho(\phi)$.

\subsection{The case when $\epsilon=-1$}

Let us first consider the case $\epsilon=-1$. In this case, the Eq.
(\ref{diff}) becomes local and can be easily solved by integration. For
symmetric $\rho(\phi)$ with zero mean, the exact solution is given by

\begin{equation}
F(x,z)=-1+ \exp\left[{z\left( {1\over {2}}-\int_0^{x}\rho(x')dx'\right)}\right],
\label{sol1}
\end{equation}
which satisfies the boundary condition $F(\infty,z)=0$ for all $z$. Expanding
the exponential in Eq. (\ref{sol1}) in powers of $z$ and using the definition in Eq. (\ref{gen}),
we find, $Q_n(x)=({1\over {2}}-\int_0^{x}\rho(x')dx')^n/n!$. Using the relation
$P_n = Q_{n+1}(-\infty)$ and the normalization condition $\int_{-\infty}^{\infty}\rho(x')dx'=1$,
we get
\begin{equation}
P_n(-1)= {1\over { (n+1)!}}\,\, ,
\label{pn1}
\end{equation}
for all $n\geq 1$. Remarkably $P_n(-1)$ is independent of the distribution $\rho(\phi)$
for {\it all} $n\geq 0$.

\subsection{The case when $\epsilon=1$}

Next we consider the case $\epsilon=1$. We first make a change of
variable, $u(x)=\int_0^{x}\rho(\phi)d\phi$. Let $F(x,z)={\tilde F}(u,z)$.
Since $\rho(\phi)$ is symmetric about zero, $u(-x)=-u(x)$ and hence
$F(-x,z)={\tilde F}(-u,z)$. Using this in Eq. (\ref{diff}) with
$\epsilon=1$, we find
\begin{equation} 
{ {\partial {\tilde F}(u,z)}\over {\partial
u}}=-z\left[1+{\tilde F}(-u, z)\right], 
\label{diff2} 
\end{equation} 
where
$u$ varies from $-1/2$ to $1/2$ and the boundary condition is, ${\tilde
F}(1/2 ,z)=0$ for all $z$. Differentiating Eq. (\ref{diff2}) with respect to $u$
we get a local second order differential equation 
\begin{equation}
{ {\partial^2 {\tilde F}(u,z)}\over {\partial u^2}} = - z^2 \left[1+{\tilde F}(u,z)\right],
\end{equation}
whose general solution is given by
\begin{equation}
{\tilde F}(u,z)= -1 + \left[ A(z)\cos (zu)+ B(z)\sin (zu)\right].
\label{sol2} 
\end{equation} 
If this solution also has to satisfy Eq. (\ref{diff2}), we have additionally $B(z)=-A(z)$.
The boundary condition ${\tilde F}(1/2,z)=0$
determines $A(z)$ and we finally get 
\begin{equation} 
F(x,z) = -1 + {{\cos \left( u(x)z\right)-\sin \left( u(x)z\right)}\over {\cos (z/2)-\sin
(z/2)}}. \label{sol3} 
\end{equation} 
Thus, $F(-\infty, z)= 2/[\cot
(z/2)-1]$. This function has poles at $z={\pi/2}+2m\pi$, where $m$ is an
integer. One can then easily evaluate the contour integration in Eq.
(\ref{cauchy}) and we get the exact expression, 
\begin{equation} P_n(1)=
2\sum_{-\infty}^{\infty} {1\over { {\left( {\pi\over
{2}}+2m\pi\right)}^{n+2}}}, \label{pn} \end{equation} valid for any $n\geq
0$. For example, by summing the series in Eq. (\ref{pn}) we find,
$P_0(1)=1$, $P_1(1)=1/2$, $P_2(1)=1/3$, $P_3(1)=5/24$, etc. which can also
be verified by performing the direct integration in Eq. (\ref{qnx}). The
remarkable fact is $P_n(1)$ is universal for {\it all} $n\geq 0$ in the
sense that it is independent of the distribution $\rho(\phi)$, as in the
$\epsilon=-1$ case. Clearly the leading asymptotic behavior is governed by
the $m=0$ term in Eq. (\ref{pn}) and we get, $P_n(1)\sim \exp(-\theta n)$
for large $n$, with $\theta(1)=\log (\pi/2)$. Clearly the exponent
$\theta(1)$ is also universal.

Interestingly, $P_n(1)$ is related to the fraction of metastable states in an Ising
spin glass on a $1$-dimensional lattice of $n$ sites at zero temperature\cite{LD}.
Consider the spin glass Hamiltonian on a chain, $H=-\sum_{i} J_{i,i+1}s_is_{i+1}$ where
$s_i=\pm 1$ are Ising variables and the bonds $J_{i,i+1}$'s are independent
and identically distributed variables each drawn from the same symmetric distribution with zero mean.
Out of the $2^n$ number of total configurations, how many are metastable with
respect to single spin flip Glauber dynamics at zero temperature? A configuration
is metastable at zero temperature if the energy change
${\Delta E}_i=2 s_i [J_{i-1,i}s_{i-1}+J_{i,i+1}s_{i+1}]\geq 0 $ due to the flip of every
spin
$s_i$. Defining the new variable $\phi_i=2 J_{i,i+1}s_{i}s_{i+1}$, we see that
the variables $\phi_i$'s are also independent and identically distributed
and the probability that a configuration is metastable is precisely
the probability that the variables, $\psi_i=\phi_i +\phi_{i-1}$ are positive
for each $i$. This is precisely $P_n(1)$ as computed in the previous paragraph.
We note that the average number
of metastable configurations $\langle N_s\rangle $ for the $1$-d spin glass was computed
exactly by Derrida and Gardner\cite{DG} by a different method and they found $\langle
N_s\rangle \sim (4/\pi)^n$
for large $n$.
Thus the fraction of metastable configurations scales as ${\langle N_s\rangle}/{2^n}\sim
(\pi/2)^{-n}$,
in agreement with our exact result for $P_n(1)$.

\subsection{The case $-1<\epsilon<1$}

We now turn to the range, $-1<\epsilon<1$. In this range, we were unable to
calculate $P_n(\epsilon)$ exactly for arbitrary distribution $\rho(\phi)$.
However progress can be made for the uniform distribution,
\begin{eqnarray}
\rho(\phi)&=&{1\over {2}}, \,\,\,\, {\rm for}\,\,\, -1\leq \phi \leq 1 \nonumber \\
&=&0, \,\,\,\,\, {\rm otherwise}.
\label{udist}
\end{eqnarray}
For this case, it follows from Eq. (\ref{diff}) that $F(x,z)$ is independent
of $x$ for $x<-1$ and hence, $F(-\infty,z)=F(-1,z)$. Similarly, $F(x,z)=0$
for all $x\geq 1$. In the range, $-1\leq x \leq 1$, we expand $F(x,z)=\sum_0^{\infty}b_m(z)x^m$
in a power series in $x$. Subsituting this series in Eq. (\ref{diff}), we get
the recursion relation, $b_m = -b_{m-1}{z(-\epsilon)^{m-1}}/{2m}$ for all $m\geq 1$.
Thus the function $F(x,z)$ can be expressed completely in terms of only $b_0(z)$ which
is then determined from the boundary condition, $F(1,z)=0$. This determines $F(x,z)$
completely in the range $-1\leq x \leq 1$ and we find, $F(x,z)=-1 + { {f(xz)}\over {f(z)}}$,
where
\begin{equation}
f(z)={ \sum_{m=0}^{\infty}{ {(-1)^{m(m+1)/2}}\over {m!}}{\left({{z}\over {2}}\right)}^m
{\epsilon}^{m(m-1)/2} } .
\label{fz}
\end{equation}
Using $F(-\infty,z)=F(-1,z)$, we finally get
\begin{equation}
F(-\infty,z)=-1 + { {f(-z)}\over {f(z)} },
\label{fxz}
\end{equation}
where $f(z)$ is given by Eq. (\ref{fz}).
We note that the series in Eq. (\ref{fz}) and hence in Eq. (\ref{fxz}) is convergent
for all $z$ as long as $-1<\epsilon \leq 1$. In fact, for $\epsilon=1$, it is
easy to see that Eq. (\ref{fxz}) gives $F(-\infty, z)= 2/[\cot (z/2)-1]$ as before.

Substituting Eq. (\ref{fxz}) in the expression
of $P_n(\epsilon)$ in Eq. (\ref{cauchy}), we find that the leading asymptotic
decay of $P_n$ for large $n$ is governed by the pole of $F(-\infty,z)$ that
is closest to the origin. From Eq. (\ref{fxz}), the poles of $F(-\infty,z)$ in
the $z$ plane are precisely the zeroes of the function $f(z)$ in Eq. (\ref{fz})
in the $z$ plane. In particular, $P_n(\epsilon)\sim {z_+}^{-n}$ for large $n$
where $z_+$ is the zero of $f(z)$ in Eq. (\ref{fz}) closest to the origin.
The persistence exponent is then simply, $\theta=\log (z_+)$. Let
us first consider a few special cases. For $\epsilon=1$, we find from Eq. (\ref{fz}),
$f(z)=\cos(z/2)-\sin(z/2)$ indicating $z_+=\pi/2$, a result we already obtained.
For $\epsilon=0$, we find from Eq. (\ref{fz}), $f(z)=1-z/2$, indicating $z_+=2$,
as expected for the persistence of uncorrelated variables. As $\epsilon\to -1^{+}$,
the function $f(z)$ is Eq. (\ref{fz}) approaches to, $f(z)\to \exp(-z)$ indicating
$z_+\to \infty$. Indeed putting $\epsilon=-1+\delta$ in Eq. (\ref{fz}), it
is easy to see that, $z_+ \approx \sqrt{8/\delta}$ as $\delta\to 0$. Thus
the persistence exponent diverges as, $\theta \approx \log [\sqrt{8/(1+\epsilon)}]$
as $\epsilon\to -1$.

For other values of $\epsilon$ in the range  $-1< \epsilon <1$, it is easy to
evaluate $z_+$ to any arbitrary accuracy from Eq. (\ref{fz}) using Mathematica.
The exponent $\theta=\log (z_+)$ for some representative values of $\epsilon$
in the range $-1<\epsilon\leq 1$ are listed in Table 1.
The exponent $\theta(\epsilon)$ increases monotonically as $\epsilon$ decreases
from $+1$ to $-1$, diverging as $\epsilon\to -1$. For $|\epsilon|>1$,
the exponent is determined from the relation, $\theta(\epsilon)=\theta(1/\epsilon)$.
Thus in the whole range, $-\infty<\epsilon<\infty$, the exponent $\theta(\epsilon)$
is a nonmonotonic function of $\epsilon$. As $\epsilon$ varies from $-\infty$
to $\infty$, $\theta(\epsilon)$ increases monotonically in the range $[-\infty,-1]$,
then decreases monotonically in $[-1,1]$ followed by a further monotonic increase
in the range $[1,\infty]$. The slowest decay of $P_n$ occurs at
$\epsilon=1$, where $\theta(\epsilon)$ is minimum and given by the universal
value, $\theta(1)=\log (\pi/2)$.

\begin{center}
\begin{tabular}{||l||l||} \hline
 $\epsilon$ & $\theta$ \\ \hline
1.0 & $\log (\pi/2)=0.4515\ldots$  \\
3/4 & $0.4690\ldots$  \\
1/2 & $0.5155\ldots$  \\
1/4 & $0.5882\ldots$  \\
0   & $\log (2)=0.6931\ldots$ \\
-1/4 & $0.8465\ldots$ \\
-1/2 & $1.0906\ldots$ \\
-3/4 & $1.5686\ldots$ \\   \hline
\end{tabular}
\end{center}

\noindent Table 1. The exponent $\theta(\epsilon)$ for some representative values of
$\epsilon$ in the range, $-1< \epsilon\leq 1$ in the case of the uniform
distribution in Eq. (\ref{udist}).
\medskip

Except at $\epsilon=0$, $1$ and $-1$, the exponent $\theta(\epsilon)$ is
nonuniversal in the sense that its value depends on the details of the
distribution $\rho(\phi)$. To see this explicitly, we now compute
$\theta(\epsilon)$ perturbatively for small $\epsilon$. We expand
the right hand side of Eq. (\ref{diff}) upto order $\epsilon$ and then solve
the resulting local differential equation exactly to determine $F(x,z)$
upto $O(\epsilon)$. Taking $x\to -\infty$ limit in the expression of $F(x,z)$, we find
\begin{equation}
F(-\infty, z)= {{2z}\over { \left[2-z-2c\rho(0)\epsilon z^2\right]}}\,\, ,
\label{pert1}
\end{equation}
where $c=\int_0^{\infty}\phi \rho(\phi) d\phi$. From Eq. (\ref{pert1}), the pole closest to
the origin is given by,
\begin{equation}
z_+= 2\left[1-4c\rho(0)\epsilon + O(\epsilon^2)\right].
\label{pole}
\end{equation}
From Eq. (\ref{cauchy}), it then follows that $P_n(\epsilon)\sim {z_+}^{-n}$ for large $n$.
Hence $\theta(\epsilon)=\log (z_+)=\log (2)- 4c\rho(0) \epsilon +O(\epsilon^2)$
and is clearly nonuniversal, as seen from the nonuniversality of the
$O(\epsilon)$ term in the previous equation. For example, for the
uniform distribution in Eq. (\ref{udist}), we get $\theta(\epsilon)=\log (2)-\epsilon/2
+O(\epsilon^2)$. On the other hand for the Gaussian distribution,
$\rho(\phi)=(2\pi)^{-1/2}\exp(-\phi^2/2)$, we get $\theta(\epsilon)=\log (2)-2\epsilon/\pi
+O(\epsilon^2)$.

\section{Conclusion}

In summary, we have discussed persistence in stochastic processes that are
stationary only integer translations of time. Such a process can be
explicitly constructed by smearing independent noises with a convolution
function $f(t)$. A physical example of such a process is provided by the
diffusion field on a hierarchical lattice for which we have computed the
smearing function $f(t)$ exactly. However, we could not compute the
persistence exponents $\theta_c$ or $\theta_d$ in this case.  We showed
that under certain conditions, the continuous-time persistence of such a
process reduces to the persistence of a discrete sequence obtained by
measuring the process only at integer times. We have constructed a
specific non-Markovian sequence where the smearing function is nonzero
only at two consecutive integer points leading to nonzero correlations
only between consecutive values of the sequence and computed the
persistence exponent $\theta_d$ exactly for this sequence. The exponent
$\theta_d$ depends continuously on the strength $\epsilon$ of the
correlation. Remarkably for $\epsilon=1$ and $\epsilon=-1$, the
persistence becomes universal. For $\epsilon=1$, we have shown an
interesting connection between the persistence of this sequence to the
average fraction of metastable states in a one dimensional spin glass.

The class of functions $f(t)$ for which we could show that $\theta_c
=\theta_d$ is perhaps not the most general. A precise characterization of
this class seems like an interesting problem. Calculation of the
persistence for SIS processes, or sequences, with correlations extending
beyond nearest neighbors may be possible in some special cases, and would
help understand the general question about the dependence of the
persistence exponent on the correlations in the sequence.

We thank A. Lef\`evre, D.S. Dean, V. S. Borkar and A.J. Bray for useful
discussions, and M. barma for a critical reading of the manuscript.

\end{multicols}

\end{document}